# Orbital angular momentum-entanglement frequency transducer

Zhi-Yuan Zhou,[1,2] Shi-Long Liu,[1,2] Yan Li,[1,2] Dong-Sheng Ding,[1,2] Wei Zhang,[1,2] Shuai Shi,[1,2] Ming-Xin Dong,[1,2] Bao-Sen Shi,[1,2,*] and Guang-Can Guo[1,2]

[1]*Key Laboratory of Quantum Information, University of Science and Technology of China, Hefei, Anhui 230026, China*

[2]*Synergetic Innovation Center of Quantum Information & Quantum Physics,*

*University of Science and Technology of China, Hefei, Anhui 230026, China*

*Corresponding author: drshi@ustc.edu.cn*

Entanglement is a vital resource for realizing many tasks such as teleportation, secure key distribution, metrology and quantum computations. To effectively build entanglement between different quantum systems and share information between them, a frequency transducer to convert between quantum states of different wavelengths while retaining its quantum features is indispensable. Information encoded in the photon's orbital angular momentum (OAM) degrees of freedom is preferred in harnessing the information-carrying capacity of a single photon because of its unlimited dimensions. A quantum transducer, which operates at wavelengths from 1558.3 nm to 525 nm for OAM qubits, OAM–polarization hybrid entangled states, and OAM entangled states, is reported for the first time. Non-classical properties and entanglements are demonstrated following the conversion process by performing quantum tomography, interference, and Bell inequality measurements. Our results demonstrate the capability to create an entanglement link between different quantum systems operating in photon's OAM degrees of freedoms, which will be of great importance in building a high capacity OAM quantum network.

Qubits and entanglement are the key resources of quantum communications and computations [1, 2]. Different physical systems such as photon pairs [2], trapped ions [3], and cold atomic gases [4] can be used to encode qubit states or generate entanglement. The photon has proved most suitable in transferring information between different systems such as quantum memory and quantum processors, or along communication channels. For a photonic qubit or an entanglement state, information can be encoded in various degrees of freedom; entanglement can be constructed via a photon's polarization [5], in time-bins [6], and in orbital angular momentum (OAM) [7]. Among these degrees of freedom of light, its OAM degrees of freedom provide unique features including the mechanical torch effect, and singularities in phase-intensity distributions, which have broad applications in micro-particle manipulations [8], high precision optical metrology [9–11] and potential high-capacity information encoding in optical communications [12, 13].

In recent years, much effort has gone into exploiting OAM light in quantum information technologies. Since the pioneering work demonstrating entanglement in OAM degrees of freedom [7], great advances have been made in the experimental control of OAM superposition states and their use in various protocols; they include quantum cryptography [14], the demonstration of very-high dimensional entanglement [15, 16], and quantum teleportation from spin to OAM degrees of freedom [17]. More recently, quantum memory for OAM qubits [18, 19] and entangled states [20–22] were demonstrated showing the capability for high-density information encoding and processing. Beyond their fundamental significance, these experiments give testimony to the potential of OAM of light as a basic information carrier and its promise in enhanced information encoding and processing capacities. A complete high capacity quantum network operating in the photonic OAM degrees of freedom should have some basic components that serve as quantum memory, quantum processors or communication channels [1]. To realize these components, the underlying physical systems usually work at different wavelengths. Transferring information between these systems effectively requires wavelength bridges which map the frequencies of one photon to another photon while preserving its quantum features.

Such wavelength bridges can be realized using second-order nonlinear processes, in which two optical fields combine in a nonlinear medium to generate a third field [23]. Energy, linear optical momentum, and OAM are conserved in the interaction process. By using high-efficiency quasi-phase-matching nonlinear waveguides, much progress in building a wavelength bridge between various systems has been made. In 2005, Tanzilli and colleagues demonstrated that time-bin entanglement between two photons at 1555 nm and 1312 nm generated by spontaneous parametric down conversion (SPDC) was preserved after up-conversion from 1312-nm to 712.4-nm [24]; in 2010, Rakher and colleagues verified that a single telecom-band photon at 1.3 μm generated from a quantum dot can be up-converted to 710 nm [25]; in 2011, Ikuta and colleagues showed down-conversion for a polarization-entangled photon at 780 nm to 1522 nm [26]; in 2014, Vollmer and collaborators demonstrated up-conversion of a 1550-nm squeezed vacuum state to 532 nm [27]. Recent progress in up-conversion detectors allows photon detection using high-performance visible optical detectors, although in each case the light detected is a highly attenuated laser source [28]. In reviewing the above work, we find that the spatial modes used are fundamental Gaussian modes because nonlinear waveguides supporting high-order spatial mode are still unobtainable. Until recently, quantum frequency conversion of OAM quantum states had been an open problem. We demonstrated the up-conversion of heralded 1560-nm single photon OAM states to 525 nm [29]. Because an OAM entanglement state has the capability to realize more sophisticated applications in quantum information science that cannot be accomplished using single-photon OAM states, frequency conversion of an OAM entangled state would be an important step in quantum information science.

In this work, we move towards this step by successively up-converting 1558.3-nm OAM qubits, OAM–polarization hybrid-entangled states, and OAM entanglement states to 525 nm. With more advanced experimental techniques compared with those used in Ref. [29], and by performing quantum state tomography, two-photon interference, and CHSH-inequality measurements, we clearly show that the quantum superposition and entanglement of the states are retained after up-conversion.

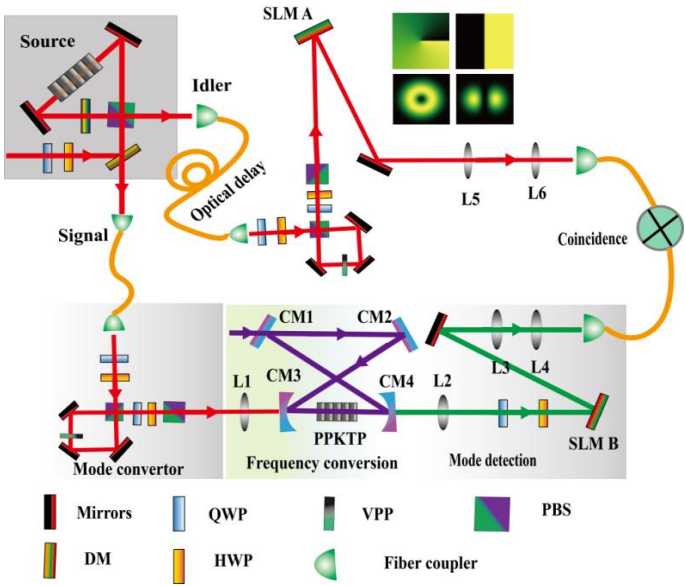

FIG. 1. Schematic of our experimental set-up. For OAM-qubit up-conversion, the Sagnac loop is operated in a single circulation direction to generate degenerate orthogonally polarized photon pairs at 1558.3 nm. The mode converter can be used to encode arbitrary qubit states or for converting entanglement type from polarization to OAM degrees of freedom. The frequency conversion is performed in a ring cavity, which is pumped at 791.0 nm; mode detection is performed using spatial light modulator (SLM) B (SLMB; HOLOEYE, LETO, 1920×1080 resolution, 6.4-μm pixel pitch). For the OAM qubit and OAM–polarization hybrid entanglement up-conversion experiments, the idler photon is not converted to an OAM mode; for OAM entanglement up-conversion experiments, the idler photon is first delayed with a single-mode fiber, and then passed through a mode converter and mode detection module. The mode detection is performed by another SLM (SLMA, 8-μm pixel pitch). Finally, the frequency up-converted signal photon and idler photon are detected by InGaAs and Si-avalanched single-photon detectors and subsequently coincidence measurements are performed (Timeharp 260, Pico Quanta, 1.6-ns coincidence window).

Frequency up-conversion of a quantum state can be accomplished using sum frequency generation (SFG), in which the annihilation of a strong pump photon ($\omega_p$) and a weak signal photon ($\omega_1$) creating a SFG photon with frequency ($\omega_2=\omega_1+\omega_p$). The effective Hamilton operator for this process is [27, 29]

$$\hat{H}_{eff} = i\hbar\xi(\hat{a}_{1,l}\hat{a}^{\dagger}_{2,l} - \hat{a}^{\dagger}_{1,l}\hat{a}_{2,l}), \tag{1}$$

where $\hat{a}_{1,l}$ and $\hat{a}^{\dagger}_{2,l}$ represent, respectively, the annihilation and creation operators of the signal and SFG photons; $l$ denotes the OAM index of the signal and SFG photons, because of OAM conservation in the SFG process, the signal photon's OAM is linearly transferred to the SFG process; $\xi$ is a constant, which is proportional to the product of the pump amplitude $E_p$ and the second-order susceptibility $\chi^{(2)}$. The evolution of $\hat{a}_{j,l}$ obtained in the Heisenberg's picture is given as:

$$\hat{a}_{1,l}(t) = \hat{a}_{1,l}(0)\cos(\xi t) - \hat{a}_{2,l}(0)\sin(\xi t), \tag{2}$$

$$\hat{a}_{2,l}(t) = \hat{a}_{2,l}(0)\cos(\xi t) + \hat{a}_{1,l}(0)\sin(\xi t). \tag{3}$$

When condition $\xi t_f = \pi/2$ holds, the input signal field is completely converted to the output SFG field $\hat{a}_{2,l}(t_f) = \hat{a}_{1,l}(0)$. As $\xi$ strongly depends on the pump amplitude, the key point for reaching maximum conversion efficiency is to increasing the pump power. In this work, the conversion efficiency is increased using a ring cavity to enhance the pump power.

Up-conversion of OAM qubit states is investigated next. The 1558.3-nm photon source used in the experiments is generated using SPDC with a type-II PPKTP crystal in a Sagnac loop configuration (See supplementary materials for details). For this up-conversion experiment, the Sagnac loop is operated in a single circulation direction by rotating the wave plates of the pump beam. The OAM qubits are generated using a modified Sagnac loop with a vortex phase plate (VPP) placed in the loop [29–31] (See mode converter in Fig. 1). The function of the mode converter is to generate OAM qubit states,

$$\alpha|l\rangle + \beta|-l\rangle, \quad (4)$$

where $|\alpha|^2 + |\beta|^2 = 1$, and $\alpha$ and $\beta$ depend on the positions of the (half, quarter) wave plates (HWP, QWP) in the input ports of the mode converter; $l$ is the photon's topological charge generated with VPP. Although $l=1$ is used in all experiments, all experimental results can be extended to other $l$ values. For convenient, we denote the qubit basis by $|R\rangle = |l\rangle$ and $|L\rangle = |-l\rangle$. We characterize the performance of our up-conversion device by converting a set of qubit states distributed over the Bloch sphere and subsequently performing quantum state tomography on the up-converted states. Reconstructing the density matrix $\hat{\rho}$ of any two-dimensional states requires the measurement of four Stokes parameters that appear in the expansion [32]

$$\hat{\rho} = \frac{1}{2}\sum_{i=0}^{3}\frac{S_i}{S_0}\hat{\sigma}_i, \quad (5)$$

where $\hat{\sigma}_0$ is the identity matrix and $\hat{\sigma}_i$ ($i=1, 2, 3$) are the Pauli spin operators. To evaluate coefficients $S_i$, projection measurements are performed on the four spatial bases (|R>, |L>, |H>, |A>) defined by $|H\rangle = 1/\sqrt{2}(|R\rangle + |L\rangle)$, $|V\rangle = 1/\sqrt{2}(|R\rangle - |L\rangle)$, $|A\rangle = 1/\sqrt{2}(|R\rangle - i|L\rangle)$, and $|D\rangle = 1/\sqrt{2}(|R\rangle + i|L\rangle)$. The projection measurements are conducted using a spatial light modulator (SLM), which is calibrated at wavelength 532 nm. By imprinting a phase mask on the SLM with opposite phase content, the spiral phase front of the SFG photon is flattened to a plane wave front that can be effectively coupled to a single-mode fiber (SMF); for details of the phase mask, see supplementary materials. Density matrices are reconstructed using the maximum likelihood method from the experimental data. Experimental reconstructed density matrices for the qubit states |R>, |L>, |H>, |D> are shown in Fig. 2; the first column shows their phase and intensity distributions and the second and third columns give the real and imaginary parts of the qubit states (for comparison, the ideal density matrices are supplied in supplementary materials, Fig. S2). In Table 1, we give the fidelities of the four qubit states without (with) dark count coincidence subtracted. The fidelity is defined as $\langle\Phi|\rho|\Phi\rangle$, where $|\Phi\rangle$ is the ideal qubit state. The average fidelity is 0.954±0.016 (0.963±0.012) without (with) the dark count subtracted. The slightly low fidelity of the |D> state is because of imperfect preparation of the input state. Nevertheless, the high fidelity shows the reliable performance of our up-conversion device, and hence paves the way for OAM entanglement states up-conversion.

| Input modes | Raw fidelity | Net fidelity |
|---|---|---|
| |R> | 0.970±0.007 | 0.977±0.005 |
| |L> | 0.978±0.004 | 0.982±0.004 |
| |H> | 0.952±0.036 | 0.967±0.023 |
| |D> | 0.916±0.015 | 0.926±0.016 |

Table 1. Fidelities of four qubits in the up-conversion process without and with dark count coincidences subtraction.

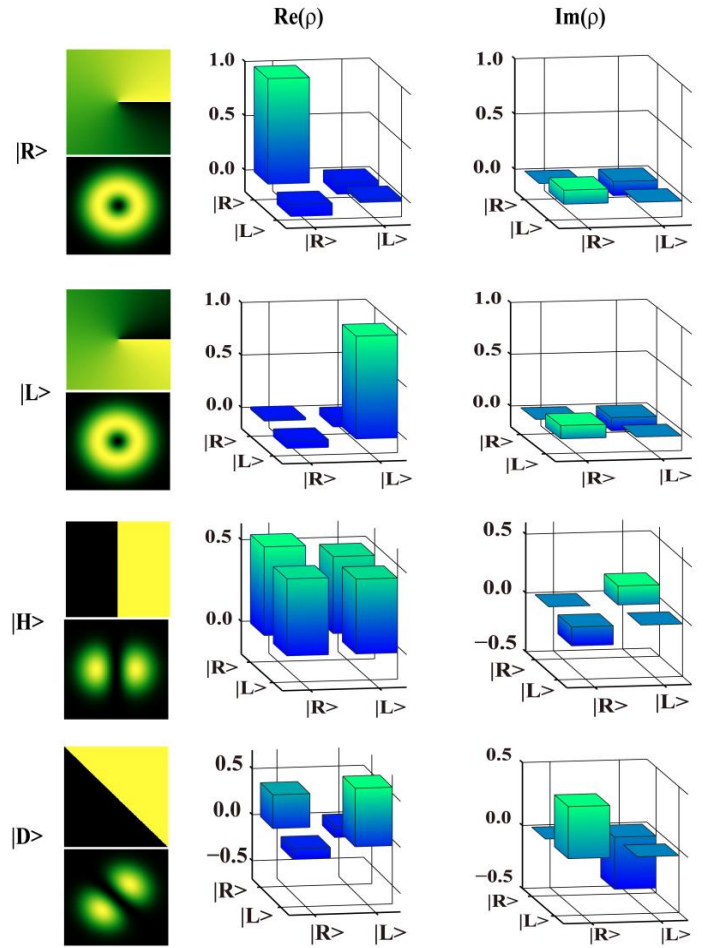

FIG. 2. Quantum tomography of up-converted OAM qubit states. Reconstructed density matrices for the four qubit states |R>, |L>, |H>, |D>. The first column shows the intensity and phase distributions imprinted on SLM A. The second and third columns give the real and imaginary parts of the density matrices. No background correction is applied.

Next, we describe up-conversion of the OAM–polarization hybrid-entangled state. Such entanglement links two different degrees of freedom of the photon. Hybrid entanglement is generated by mode conversion from a Sagnac-loop-based polarization entangled source, which can generate entangled

states [33]

$$\Phi^\pm = \frac{1}{\sqrt{2}}(|hv\rangle \pm |vh\rangle). \quad (6)$$

By performing mode conversion on one of the photons, the state is transformed to an OAM–polarization hybrid-entangled state of the form [31]

$$|\Phi\rangle_{hybrid}^\pm = \frac{1}{\sqrt{2}}(|h,R\rangle \pm |v,L\rangle). \quad (7)$$

In this experiment, the state $|\Phi\rangle_{hybrid}^+$ is used. The 1558.3-nm signal photon in the OAM mode is sent to the frequency conversion module, which is then up-converted to 525 nm. It is subsequently transformed to a Gaussian mode using the mode-detection module and coupled to a SMF. The idler photon is optically delayed with a SMF and together with the up-converted signal photon coincidence measurements are made.

To verify that entanglement is preserved during up-conversion for the hybrid-entangled state, two-photon interference and quantum state tomography are used to characterize the up-converted state. Two interference fringes are measured when the idler photon is polarized in the diagonal (|d>) or right circular (|r>) state. For each polarization setting, we record coincidences over a 100-s period as a function of the rotation angle of the phase mask applied to SLM B. The definition of the angle is $|\theta\rangle = \frac{1}{\sqrt{2}}(e^{i\theta}|R\rangle + e^{-i\theta}|L\rangle)$; for details see supplementary materials Fig. S3. The interference visibilities without (with) dark count coincidences subtracted are 0.949±0.029 (0.972±0.027) and 0.856±0.068 (0.907±0.053), for |d> and |r>, respectively.

For OAM–polarization hybrid entanglement, survival of entanglement after conversion can also be characterized by entanglement witness, which is a commonly used method to infer the non-classical correlation of the state. Entanglement witness is defined as $W = V_{d/a} + V_{r/l}$ with visibilities $V_{d/a}$ and $V_{r/l}$ for the different polarization states (d=diagonal, a=anti-diagonal, r=right circular, l=left circular). For a separable state, $W \leq 1$ [31]; in our experiments, the calculated witness value is 1.805±0.097 (1.879±0.080) without (with) dark count coincidence subtracted, which violates the inequality by more than eight standard deviations. To know precisely what the up-converted state is, one needs to perform quantum state tomography to reconstruct the density matrix of the state; the result is shown in Fig. 3(c) and (d). By comparing with the ideal hybrid state $|\Phi\rangle_{hybrid}^+$, the fidelity of the up-conversion $_{hybrid}\langle\Phi^+|\rho|\Phi^+\rangle_{hybrid}$ is 0.837±0.025 (0.861±0.036) without (with) dark count coincidence subtracted. The data acquisition time for quantum state tomography measurements is 200 s. The uncertainty errors are estimated by assuming a Poisson distribution for the photon statistics.

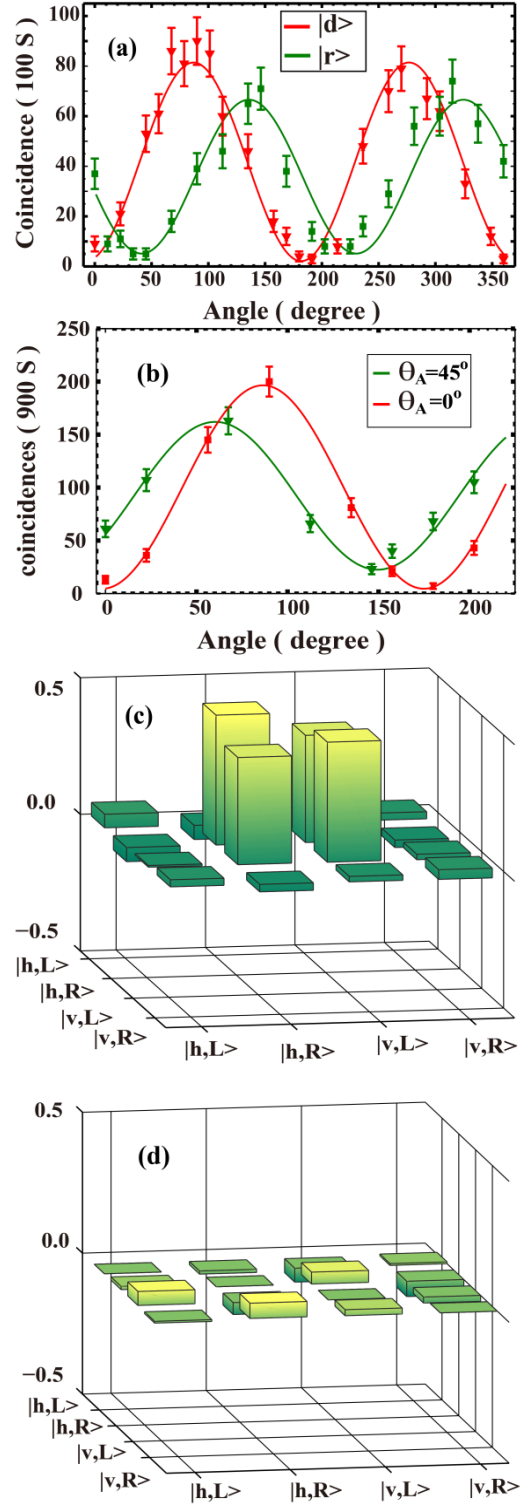

FIG. 3. Two-photon interference fringes and quantum state tomography for the OAM–polarization hybrid-entangled state and the OAM entangled state. (a) Coincidence counts over a 100-s period as a function of the rotation angle of phase mask applied to SLM B when the polarization of the idler photon polarization remains diagonal (|d>) and right-circular polarized (|r>). (b) Coincidences over a 900-s period as function of the rotation angle of the phase mask in SLM B when the angle of SLM A is fixed at 45° and 0°.

Finally, we describe the up-conversion of an OAM entangled state. When the idler photon is also transformed with the mode converter, the polarization entangled state is transformed to an OAM entangled state [30]

$$\Phi_{\text{OAM}}^{\pm} = \frac{1}{\sqrt{2}}\left(|R,L\rangle \pm |L,R\rangle\right). \quad (8)$$

For this up-conversion experiment, the state $|\Phi\rangle_{\text{OAM}}^{-}$ is used. The signal photon undergoes the same procedures as described in the previous section. The idler photon is passed through a delay SMF, mode converter, and mode detection module. Finally, a coincidence measurement between the idler photon and the up-converted signal photon is performed. To demonstrate entanglement is preserved during up-conversion, we measured the two-photon interference fringes first [Fig. 3(b)]. Coincidences over a 900-s period are recorded as a function of the rotation angle of the phase mask applied to SLM B when the angle of the phase mask in SLM A is set at 0 ° and 45 °. The visibilities for the 0 ° and 45 ° bases without (with) dark count coincidences subtracted are 0.955±0.023 (0.994±0.008) and 0.750±0.062 (0.784±0.059), respectively. Visibilities greater than 71% indicate possible violation of the Bell inequality, which implies the presence of entanglement. To further characterize the entanglement property of the up-converted state, we check the *S* parameter of the CHSH inequality defined as [34],

$$S = E(\theta_A, \theta_B) - E(\theta_A, \theta_B') + E(\theta_A', \theta_B) + E(\theta_A', \theta_B'), \quad (9)$$

where $E(\theta_A, \theta_B)$ is expressed as

$$E(\theta_A,\theta_B) = \frac{C(\theta_A,\theta_B) + C(\theta_A+\pi/2,\theta_B+\pi/2) - C(\theta_A+\pi/2,\theta_B) - C(\theta_A,\theta_B+\pi/2)}{C(\theta_A,\theta_B) + C(\theta_A+\pi/2,\theta_B+\pi/2) + C(\theta_A+\pi/2,\theta_B) + C(\theta_A,\theta_B+\pi/2)}. \quad (10)$$

In our experiment, the settings for the four angles are: $\theta_A = 0$ and $\theta_B = \pi/8$; $\theta_A' = \pi/4$ and $\theta_B' = 3\pi/8$. For classical correlations, $|S| \leq 2$. Our measurements give a *S* value of 2.39±0.12 (2.50±0.09) without (with) dark count coincidence subtracted, which violates the inequality by more than three standard deviations. Thus, we have strong evidence for the presence of entanglement after frequency up-conversion.

Up to this point, we have described a quantum frequency transducer for OAM qubits, and OAM–polarization hybrid-entangled states and OAM-entangled states enabling conversions from 1558.3 nm to 525 nm. Our results answer basic questions concerning quantum frequency conversion of OAM states. Is it possible to convert OAM entanglement from one wavelength to another wavelength? Do superpositions of OAM states and entanglements survive after frequency conversion? Our demonstrations give positive answers to these questions. Nevertheless, there are other important issues that need to be solved. One is the quantum efficiency during conversion, another is how to increase the dimensions of the states for conversion. Feasible resolutions of these issues are given in the following.

In theory, there is no limitation in increasing the conversion efficiency to unity. For increasing the overall quantum conversion efficiencies, first the bandwidth of the photon source for conversion should match the bandwidth of the SFG crystal. In the present experiments, the quantum efficiency is 0.01 for a coherent narrow bandwidth laser, and is 0.002 for the signal photon because only 20% of the photons are in the effective bandwidth of the SFG crystal. This problem can be solved by using a longer crystal in SPDC to generating photon pairs and a shorter crystal for frequency up-conversion. The key point in increasing conversion efficiency is to increase the pump power. For up-conversion of continuous-wave photons, one needs to optimize the cavity loss and the transmittance of the input coupling mirror; for this a high-intensity pump laser is needed. For pulsed photons, such a laser is preferred in achieving high conversion efficiency. Researchers already have realized high conversion efficiency for lower-order OAM modes in the pulsed regime using an attenuated laser source [35].

Regarding dimensions in frequency conversion, the physical limitations of the dimensions are crystal thickness and beam waist of the strong pump beam. The thickness of the crystal used in this experiment is 1 mm, as the OAM beam size $w_l$ scales with $l$ in $w_l = \sqrt{l+1}w_0$, where $w_0$ is the beam waist of the Gaussian beam. The crystal can support up to 1200 modes for $w_0$=20 μm. If the pump beam waist is limited to 100 μm, at least 49 OAM modes are effectively overlapped with the pump beams. In the classical optical regime, $l$=100 is obtained in second harmonic generation [36]. In our experiments, the $l$=1 mode is used only as an example. If the conversion efficiency issue is solved, dimensional limitations on $l$ pose no problem.

In conclusion, the experiments described above provide a first look at frequency up-conversions of a OAM qubit, OAM–polarization hybrid-entangled state, and OAM-entangled state. Various measurements including quantum state tomography, interference, and the *S*-parameter of the CHSH inequality were used to characterize the performance of the frequency converter for the various quantum states. Preservation of quantum superposition and entanglement survival demonstrate the quantum nature of the conversion. The results open doors for new research into quantum frequency conversion in the photon's OAM degrees of freedom, which will stimulate broad interest in solving the remaining issues discussed above. Up-conversion of the OAM degrees of freedom of a photon enables a quantum wavelength bridge that links two quantum systems that can be exploited in future high-capacity quantum networks.

This work was supported by the National Fundamental Research Program of China (Grant No. 2011CBA00200), the National Natural Science Foundation of China (Grant Nos. 11174271, 61275115, 61435011, and 61525504) and the Fundamental Research Funds for the Central Universities.

## Supplementary of materials

**The polarization entangled photon source.** Polarization entangled source used in our experiments is generated from type-II PPKTP in a Sagnac-loop configuration. The source is of high brightness and compact and high entanglement quality. The detail performance of the source can be found in [33]. In the present experiments, the pump laser of the source is from a self-building SHG laser. The pump wavelength is 778.15 nm, and the pump power is fixed at 150 mW. The source emitted degenerate photon pairs at 1558.3nm, the bandwidth of the source is 2.5 nm. The 1558.3 nm photon is detected using a free running InGaAs avalanched photon detector (ID220, 20% efficiency, 8 μs dead time).

**Details of the frequency conversion module.** The frequency conversion module is based on sum frequency generation (SFG) in a ring cavity. The cavity is designed for singlely resonant for strong pump beam at 791 nm. The pump power is fixed at 700 mW in the experiments. The single photon at 1558.3 nm only single passes the cavity mirrors CM3 and CM4. The ring cavity has two flat mirrors (CM1, CM2) and two concave mirrors (CM3, CM4) with curvature of 80 mm. The input mirror CM1 has 3% transmittance for 791 nm; CM2 is high reflective-coated at 791 nm (R>99.9%); mirror CM3 is high transmitted-coated at 1558.3 nm (T>98%) and high reflective-coated at 791 nm; mirror CM4 is high-reflective-coated at 791 nm, high transmitted-coated at 1558.3 nm and 525 nm (T>%98). The total length of the cavity is 430 mm and the beam waist 40 μm at the center of the PPKTP crystal. The PPKTP crystal has dimensions of 1mm×2 mm×20 mm. The phase matching temperature is kept at 51℃.

**Mode detection module.** For detecting the up-converted signal photon at 525 nm. A 150 mm lens L2 is used to image the up-converted photon to SLMB with a magnification of 20. The beam diameter at SLMB is near 2 mm. The wave front is transformed by applying a phase mask to the SLMB, then we use two lenses to coupling the transformed beam to single mode fiber. For OAM entanglement up-conversion, the idler photon is also mode detected with the same procedure by using SLMA.

**Efficiencies in up-conversion experiments.** The average collection efficiency of signal and idler photons of the entangled source is 0.26. For OAM qubit up-conversion experiments, the efficiency includes the mode conversion and free space transmission is 0.80 for the signal photon; the conversion efficiency for coherent narrow band laser beam is 0.01, the bandwidth of the SFG is 0.5 nm, which reduced the total quantum efficiency to 0.002; after up-conversion the mode detection efficiency is around 0.48, which include the mode transformation efficiency of the SLMB (0.80) and the fiber coupling efficiency (0.60). The single photon detection efficiency is 0.50. The overall efficiency of the signal photon is $1.0\times10^{-4}$. For OAM entanglements up-conversion, the idler photon's mode conversion efficiency is 0.80; only half of the photon is survived after eliminate the discrimination in photon's polarization; total efficiency of the mode convertor is 0.40; the mode detection efficiency is 0.50, which include mode transformation efficiency of SLMA (0.80) and fiber coupling efficiency (0.62); the infrared single photon detector's detection efficiency is 0.20. The overall efficiency of the idler photon is 0.01.

**How we perform OAM qubit tomography**

The OAM qubit tomography is performed by projection measurements on 4 basis |R>, |L>, |H>, |A>. Phase masks that applied to SLMB are showed in figure S1. The four projection basis can be expressed in vector form as $\chi_1 = \begin{pmatrix}1\\0\end{pmatrix}$, $\chi_2 = \begin{pmatrix}0\\1\end{pmatrix}$, $\chi_3 = \frac{1}{\sqrt{2}}\begin{pmatrix}1\\1\end{pmatrix}$, and $\chi_4 = \frac{1}{\sqrt{2}}\begin{pmatrix}1\\-i\end{pmatrix}$, respectively. By defining $\alpha = \begin{pmatrix}t_1 & 0\\ t_3+t_4 i & t_2\end{pmatrix}$, $\beta = \begin{pmatrix}t_1 & t_3-it_4\\ 0 & t_2\end{pmatrix}$, $\rho_e = \frac{\alpha\cdot\beta}{tr(\alpha\beta)}$, probabilities of projection into these four basis can be expressed as $p_i = \chi_i^\dagger \rho_e \chi_i (i=1,2,3,4)$. Assuming the coincidences in the four basis measured are $n_i (i=1,2,3,4)$, then define $N = n_1 + n_2$ as the normalized constant. By minimizing the function $L = \sum_{i=1}^{4}\frac{(Np_i - n_i)^2}{2Np_i}$, we can obtain the optimized density matrix for a qubit state.

For the qubit states |R>, |L>, |H>, |D> we generate in the experiment, the idea density matrix should be $\rho_R = \begin{pmatrix}1 & 0\\0 & 0\end{pmatrix}$, $\rho_L = \begin{pmatrix}0 & 0\\0 & 1\end{pmatrix}$, $\rho_H = \frac{1}{2}\begin{pmatrix}1 & 1\\1 & 1\end{pmatrix}$, $\rho_D = \frac{1}{2}\begin{pmatrix}1 & -i\\i & 1\end{pmatrix}$, plotting of these ideal density matrices are showed in figure S2.

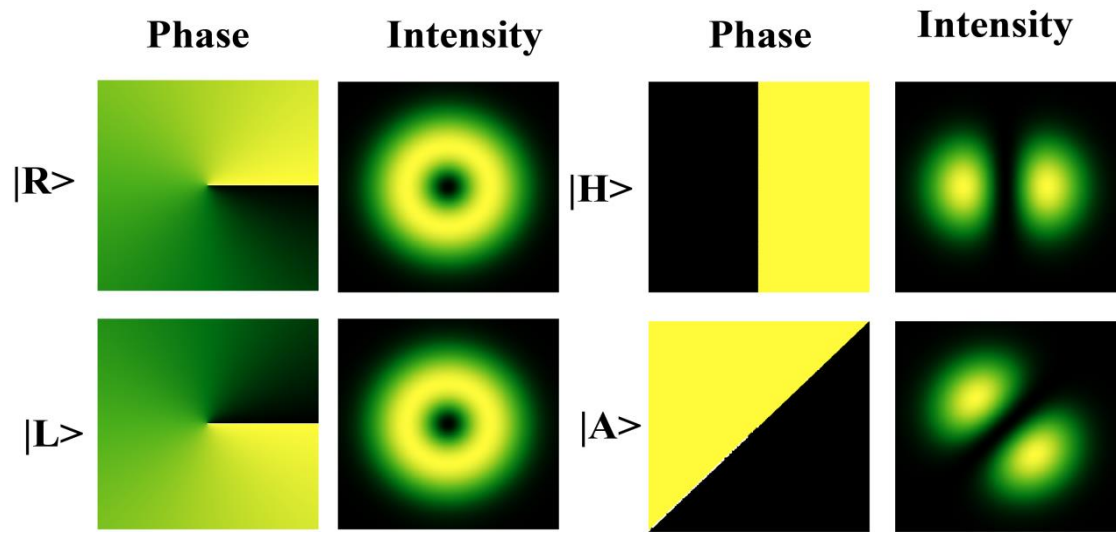

Figure S1. Phase and intensity distribution of the four projection basis used in our experiments.

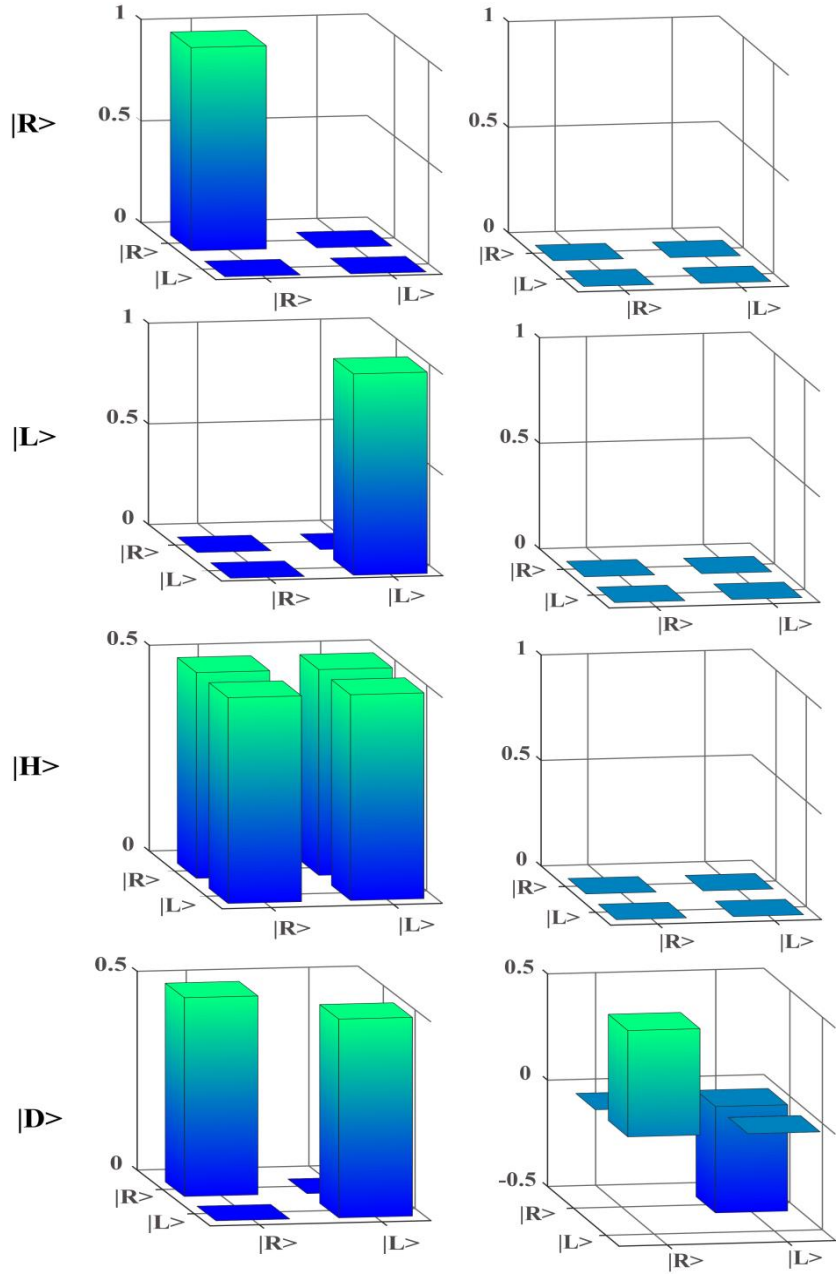

Figure S2. Real and imaginary part of the ideal density matrices for qubit states |R>, |L>, |H>, |D>.

**Definitions of the rotation angles of the phase masks in our experiments.**
In both OAM-polarization hybrid entanglement up-conversion and OAM entanglement up-conversion, we need to measure coincidence count as function of the rotation angle of the phase mask applied to the SLM, the mathematical definition of the rotation angle is expressed as $|\theta\rangle = \frac{1}{\sqrt{2}}(e^{i\theta}|R\rangle + e^{-i\theta}|L\rangle)$, the relative phase between the two superposition state is $2\theta$, figure S3 shows the diagram of the definition of the rotation angle $\theta$.

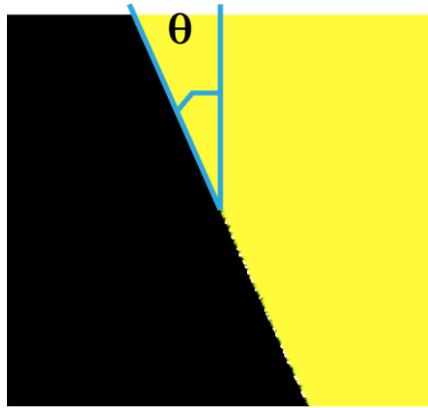

Figure S3. Diagram for the definition of the rotation angle of the phase mask applied to SLM A and B.